\documentclass[prb,aps,amsmath,twocolumn,showpacs,superscriptaddress,
longbibliography]{revtex4-1}

\usepackage{amssymb}
\usepackage{graphicx}
\usepackage[pdftex,bookmarks,colorlinks,breaklinks]{hyperref}
\hypersetup{linkcolor=red,citecolor=blue,filecolor=dullmagenta,urlcolor=blue}
\usepackage{color}

\usepackage{soul}
\definecolor{indiagreen}{rgb}{0.07, 0.53, 0.03}

\usepackage{amsmath}
\usepackage{mathrsfs}
\usepackage{amsfonts}

\begin{document}

\title{Phenomenological approach to transport through three-terminal disordered wires}

\author{A.~M. Mart\'inez-Arg\"uello}
\email{blitzkriegheinkel@gmail.com}
\affiliation{Instituto de F\'isica, Benem\'erita Universidad Aut\'onoma de 
Puebla, \\ Apartado Postal J-48, 72570 Puebla, Mexico}

\author{J.~A. M\'endez-Berm\'udez}
\email{jmendez@ifuap.buap.mx}
\affiliation{Instituto de F\'isica, Benem\'erita Universidad Aut\'onoma de 
Puebla, \\ Apartado Postal J-48, 72570 Puebla, Mexico}

\author{M. Mart\'{\i}nez-Mares}
\email{moi@xanum.uam.mx}
\affiliation{Departamento de F\'{\i}sica, Universidad Aut\'onoma
Metropolitana-Iztapalapa, Apartado Postal 55-534, 09340 Ciudad de M\'exico, Mexico}


\begin{abstract}
We study the voltage drop along three-terminal disordered wires in all transport 
regimes, from the ballistic to the localized regime. This is performed by 
measuring the voltage drop on one side of a one-dimensional disordered wire in a 
three-terminal set-up as a function of disorder. Two models of disorder in the 
wire are considered: (i) the one-dimensional Anderson model with diagonal disorder 
and (ii) finite-width bulk-disordered waveguides. Based on the known 
$\beta$-dependence of the voltage drop distribution of the three-terminal 
chaotic case, being $\beta$ the Dyson symmetry index ($\beta=1$, 2, and 4 for 
orthogonal, unitary, and symplectic symmetries, respectively), the analysis is 
extended to a continuous parameter $\beta>0$ and use the corresponding 
expression as a phenomenological one to reach the disordered phase. We show that 
our proposal encompasses all the transport regimes with $\beta$ depending linearly on the 
disorder strength.
\end{abstract}

\pacs{73.23.-b, 73.21.Hb, 72.10.-d, 72.15.Rn}

\maketitle


\section{Introduction}

Quantum transport through multiprobe mesoscopic systems and nanostructures with 
complex dynamics has been of great interest for a long time (see for 
instance 
Refs.~[\onlinecite{Buttiker1986,ButtikerIBM,Song1998,Datta,Goodnick,GodoyEPL,
Godoy,GMM1,Gao,Arrachea1,Arrachea2}] and references therein). The earlier 
experiments considered conductors of normal metal whose size is larger than the 
elastic mean free path. Quantum coherence along the sample with randomly 
distributed impurities gives rise to striking quantum interference effects, as 
well as to sample-to-sample fluctuations in the transport properties, due to the 
different microscopic configurations of disorder, that were the subject of 
intense research like the magnetoresistance, the Hall effect, persistent 
currents, among others.\cite{Webb1,Paalanena,Saeed,Levy} More recently, the 
statistical fluctuations of the transport properties through clean quantum 
devices with chaotic classical dynamics, have been 
investigated.\cite{Marcus1992,Beenakker1997,Alhassid2000,Mello2008}

Of particular interest are three-terminal systems since they offer potential 
applications;\cite{Jacobsen,Andrew2008} for instance, three terminal systems 
are used to sense the coupling strength between individual leads and the 
different modes in the device they are coupled to.\cite{ArnhildNJP} The fluctuations 
of the voltage drop along an electronic device was first studied in disordered 
wires,\cite{GodoyEPL,Godoy} while in chaotic devices was considered in 
Ref.~[\onlinecite{AngelAIP}], using random matrix theory simulations. For the 
particular configuration of the three-terminals, where the voltage probe is on 
one side of the chaotic wire, an analytical expression for all symmetry classes 
(orthogonal, unitary and symplectic) as well as an auxiliary experiment with 
chaotic microwave graphs that verifies the theoretical prediction, were 
presented in Ref.~[\onlinecite{Angel2018}].

In this paper we study the voltage drop on one side of disordered wires for all 
the transport regimes. The system is studied by the scattering matrix 
approach and, in order to validate our results, we appeal to two models to 
describe the disordered wire: The finite size one-dimensional Anderson model 
with diagonal disorder and finite-width bulk-disordered waveguides. Our 
analysis is based on the distribution of the voltage drop, whose dependence on 
the Dyson parameter $\beta$ is explicit ($\beta=1$ for the orthogonal symmetry, 
$\beta=2$ for the unitary one, and $\beta=4$ for the symplectic symmetry). 
This distribution is extended to continuous $\beta$, which is used as a 
phenomenological expression. We show that this procedure describes all transport 
regimes, deep from the ballistic to the localized regime, where the Dyson 
parameter $\beta$ may be interpreted as the degree of disorder since it depends 
only on the ratio between the localization length and the system size, which is 
a measure of disorder. Our results are in agreement with numerical simulations 
and may be verified experimentally in single-mode waveguides with either bulk or 
surface disorder.

The paper is organized as follows. In Sect.~\ref{sec:LB-approach} we summarize 
the main results about the voltage drop in three-terminal devices when the 
voltage probe is on one side of a horizontal disordered wire. Also, there, we
present the corresponding statistical distribution and emphasize the Dyson 
parameter dependence when the wire is a chaotic cavity. In 
Sect.~\ref{sec:3T1DAM} we present the description of the disordered wire in 
terms of the open one-dimensional Anderson model, while a finite-width 
bulk-disordered waveguide realization is presented in Sect.~\ref{sec:3TMS}. We 
present our conclusions in Sect.~{\ref{sec:conclusions}}.


\section{Voltage drop in a three-terminal device}
\label{sec:LB-approach}

\begin{figure}
\begin{center}
\includegraphics[width=7.0cm]{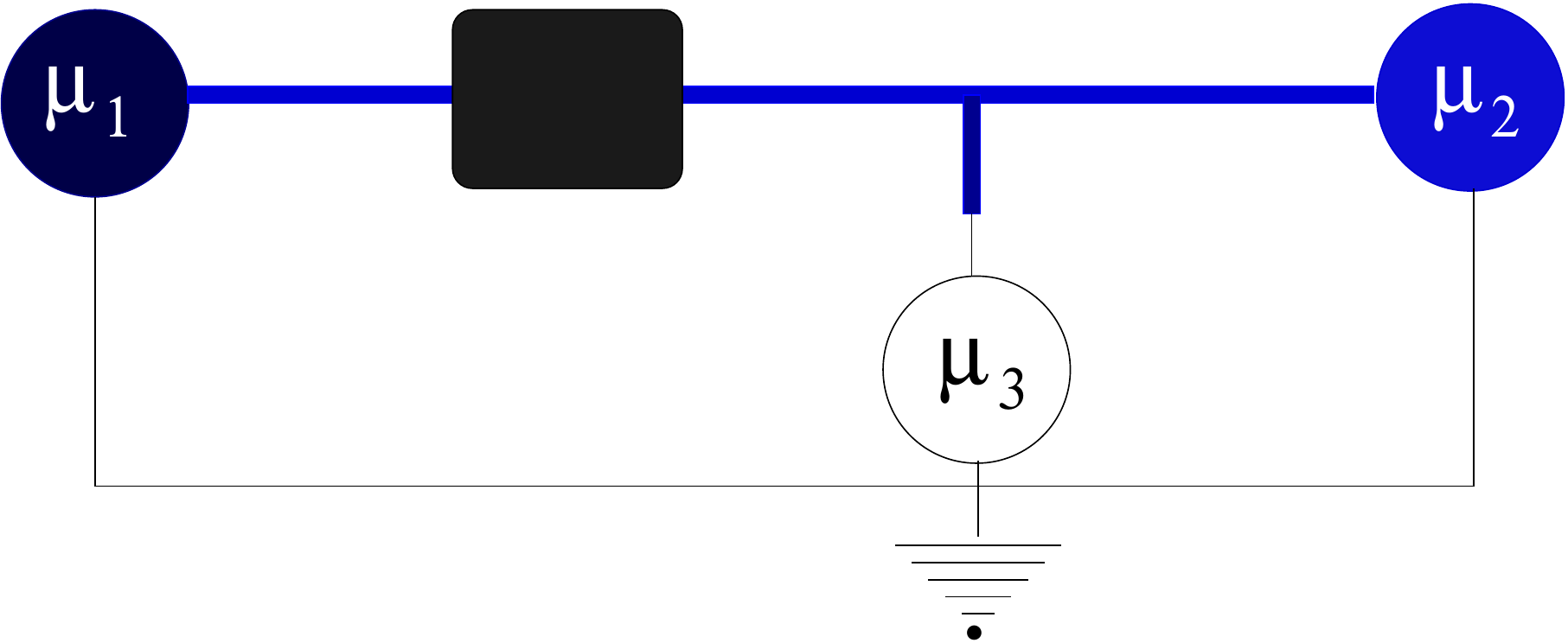}
\caption{{\footnotesize Sketch of a three-probe setting that allows the 
measurement of the voltage drop along a device, represented by the horizontal 
wire. A flux current is established along the horizontal wire, while the 
vertical wire measures the voltage drop $\mu_3$, which depends on the chemical 
potentials $\mu_{1}$ and $\mu_{2}$. The thin blue lines represent perfect 
conductors connected to the sources of voltages.}}
\label{fig:3TFig}
\end{center}
\end{figure}

In the Landauer-B\"uttiker formalism of multi-terminal devices the electronic 
transport is reduced to a scattering problem.\cite{Buttiker1986} The simplest 
arrangement that allows the measurement of the voltage drop along a device is a 
three-probe setting. As an example we consider the system shown in 
Fig.~\ref{fig:3TFig} in which the device, represented by the black box, is 
connected via perfect leads (blue lines) to fixed sources of voltages $\mu_{1} 
(= eV_{1})$ and $\mu_{2} (= eV_{2})$ that induce a flux current along the wire. 
The voltage drop can be measured by means of a third wire (vertical blue line) 
used as a voltage probe. This can be achieved by fulfilling the requirement that 
the current passing through the probe vanishes, thus yielding to the voltage 
drop $\mu_{3} (= eV_{3})$ along the device, namely~\cite{ButtikerIBM}
\begin{equation}
\mu_3 = \frac{1}{2}\left( \mu_1 + \mu_2\right) + \frac{1}{2} \left( \mu_1- 
\mu_2 \right)\, f, 
\label{eq:mu3}
\end{equation}
with
\begin{equation}
f = \frac{T_{31} - T_{32}}{T_{31} + T_{32}} ,
\label{eq:f}
\end{equation}
where $T_{31}$ and $T_{32}$ are the partial transmission probabilities 
from wire 1 to wire 3 and from wire 2 to wire 3, respectively. 

Since the electrons travel freely through each perfect lead and suffer a 
scattering process due to the disordered wire, the quantity $f$ depends on the 
intrinsic nature of the conductor and contains all the relevant information 
about the multiple scattering in the device. If the device is a disordered or a 
chaotic wire, $f$ fluctuates in the interval $[-1,1]$ since $\mu_{3}$ can not 
reach neither the value $\mu_{1}$ nor $\mu_{2}$ due to the contact 
resistance.\cite{ButtikerIBM}

We are interested in the situation in which the perfect leads are single-mode 
waveguides and that the probe is symmetrically coupled to the other two 
terminals at the junction. In that case the scattering matrix that describes the 
wire is a $2\times 2$ matrix which has the general form
\begin{equation}
\label{eq:Sw}
S =
\left( \begin{array}{cc}
r & t' \\
t & r' \\
\end{array} \right),
\end{equation}
where $r$ ($r'$) and $t$ ($t'$) are the reflexion and transmission amplitudes when 
incidence is from the left (right) of the wire, and Eq.~(\ref{eq:f}) takes the 
form\cite{Angel2018}
\begin{equation}
\label{eq:f3}
f = \frac{|t|^{2} - |1-r'|^{2}}{|t|^{2} + |1-r'|^{2}}.
\end{equation}

\subsection{Chaotic wire}

For the case in which the wire is a chaotic cavity, $S$ is chosen from an 
appropriate ensemble of scattering matrices according to the symmetry present in 
the system. That is, $S$ belongs to one of the so-called circular ensembles 
from random matrix theory, with $\beta$ representing the symmetry class 
present in the system: In the absence of any symmetry, flux conservation 
condition is the only requirement $S$ must fulfill, it becomes a unitary 
matrix, $SS^{\dagger}=\openone$ with $\openone$ the $2\times 2$ unit matrix,  
and $S$ belongs to the Circular Unitary Ensemble (CUE). The presence of time 
reversal symmetry defines the Circular Orthogonal Ensemble (COE), in which case 
$S$ is a symmetric unitary matrix, $S=S^{T}$, where $T$ stands for the 
transpose. Finally, the presence of time reversal and spin-rotation symmetries 
define the Circular Symplectic Ensemble (CSE), in which $S$ is a self-dual 
quaternion matrix and satisfies $S^{\mathrm{R}}=S$ and the flux conservation 
condition reads $SS^{*}=\openone$, where $S^{*}$ is the complex quaternion of 
$S$. In the Dyson scheme, these ensembles are labeled by $\beta=1$, 2, and 4, 
respectively.\cite{Dyson1962} For these symmetry classes, the statistical 
distribution of $f$ is given by\cite{Angel2018}
\begin{equation}
\label{eq:pfB}
p_{\beta}(f)
= \left\{
\begin{array}{l}
\frac{1}{\pi} \sqrt{(1-f)/(1+f)}  \, \, \, \, \, \mbox{for}\, {\beta=1}, \\ \\
\frac{1}{2} (1-f) \, \, \, \, \, \, \, \mbox{for}\, {\beta=2}, \\ \\
\frac{3}{4} (1+f) (1-f)^{2} \, \, \, \mbox{for}\, {\beta=4} ,
\end{array} \right.
\end{equation}
which can be written in a single equation as 
\begin{equation}
\label{eq:pfAll}
p_{\beta}(f) = 2^{1 - \beta} 
\frac{\Gamma(\beta)}{[\Gamma(\beta/2)]^{2}} 
\frac{(1 - f)^{\beta/2}}{(1 + f)^{1 - \beta/2}} .
\end{equation}

This distribution is the main quantity on which this paper is focused. We 
propose $p_{\beta}(f)$ of Eq.~(\ref{eq:pfAll}) as a phenomenological expression
where the index $\beta$ is extended to a continuous parameter that allows to 
cover all transport regimes, from the ballistic to deep in the localized regime.
To verify the validity of our assertion we make use of two models for the 
description of the disorder in the wire: The open one-dimensional Anderson model 
and finite-width bulk-disordered waveguides.


\section{Open 1D Anderson model: effective Hamiltonian approach}
\label{sec:3T1DAM}

A model of disorder in the wire can be implemented in an $N$-site 
one-dimensional wire of length $L$ described by the tight-binding Hamiltonian 
$H$ with nearest neighbor interactions of the form 
\begin{equation}
\label{eq:1DAH}
H_{mn} = \varepsilon_{n} \delta_{mn} - \nu (\delta_{m, n+1} + \delta_{m, n-1}),
\end{equation}
where $\varepsilon_{n}$ is the energy of site $n$, $\nu$ is the tunnel 
transition amplitude to nearest neighbor sites, and $\delta$ is the usual 
Kronecker delta. For diagonal disorder $\nu$ is just a constant, that we  
fix to $\nu=1$, while the site energy $\varepsilon_{n}$ is a random number 
which for simplicity we consider uniformly distributed in the interval 
$[-w/2,w/2]$ with variance $\sigma^2=\langle\varepsilon^2_n\rangle=w^2/12$, 
being $w$ a measure of the amount of disorder.

We open the wire by attaching it on the left ($L=1$) and right ($L=N$) ends to 
semi-infinite single-mode perfect leads with coupling strength 
$\gamma^{\mathrm{L,R}}$ to the left ($\mathrm{L}$) and to the right  
($\mathrm{R}$) end, respectively. The $2\times 2$ $S$-matrix can be 
written in the form\cite{Datta}
\begin{equation}
\label{eq:S1DAM}
S(E) = \openone - 2\mathrm{i} \sin(k) W^{T} 
\frac{1}{E - \mathcal{H}_{\mathrm{eff}} } W
\end{equation}
where $E$ is the energy, $k=\arccos(E/2)$ is the wave vector supported in the 
leads, and $\mathcal{H}_{\mathrm{eff}}$ is the effective non-Hermitian 
Hamiltonian, namely
\begin{equation}
\label{eq:Heff}
\mathcal{H}_{\mathrm{eff}} = H - \frac{ \mathrm{e}^{\mathrm{i} k} }{2} W W^{T}.
\end{equation}
In Equations~(\ref{eq:S1DAM}) and (\ref{eq:Heff}) the matrix $W(E)$ describes 
the coupling of the wire with the leads. Its elements are defined by 
\begin{equation}
W_{mn} = 2\pi \sum_{c=\mathrm{L,R}} A^{c}_{m}(E) A^{c}_{n}(E),
\end{equation}
with the coupling amplitudes
\begin{equation}
A^{\mathrm{L,R}}_{n}(E) = \sqrt{ \frac{ \gamma^{\mathrm{L,R}} }{\pi} } \left( 1 - \frac{E^{2}}{4} \right)^{1/4} \left(\delta^{\mathrm{L}}_{n,1} + \delta^{\mathrm{R}}_{n,N} \right).
\end{equation}
Furthermore, the energy dependence in $\mathcal{H}_{\mathrm{eff}}$ can be 
neglected since $\arccos(E/2)$ changes slightly at the center of the band. 
Moreover, the inverse localization length reduces to 
$\ell^{-1}_{\infty}(E)=w^{2}/105.2$,\cite{Kappus1981} which means that the higher 
the intensity of disorder the smaller the localization length is, as expected.

\begin{figure}
\begin{center}
\includegraphics[width=8.5cm]{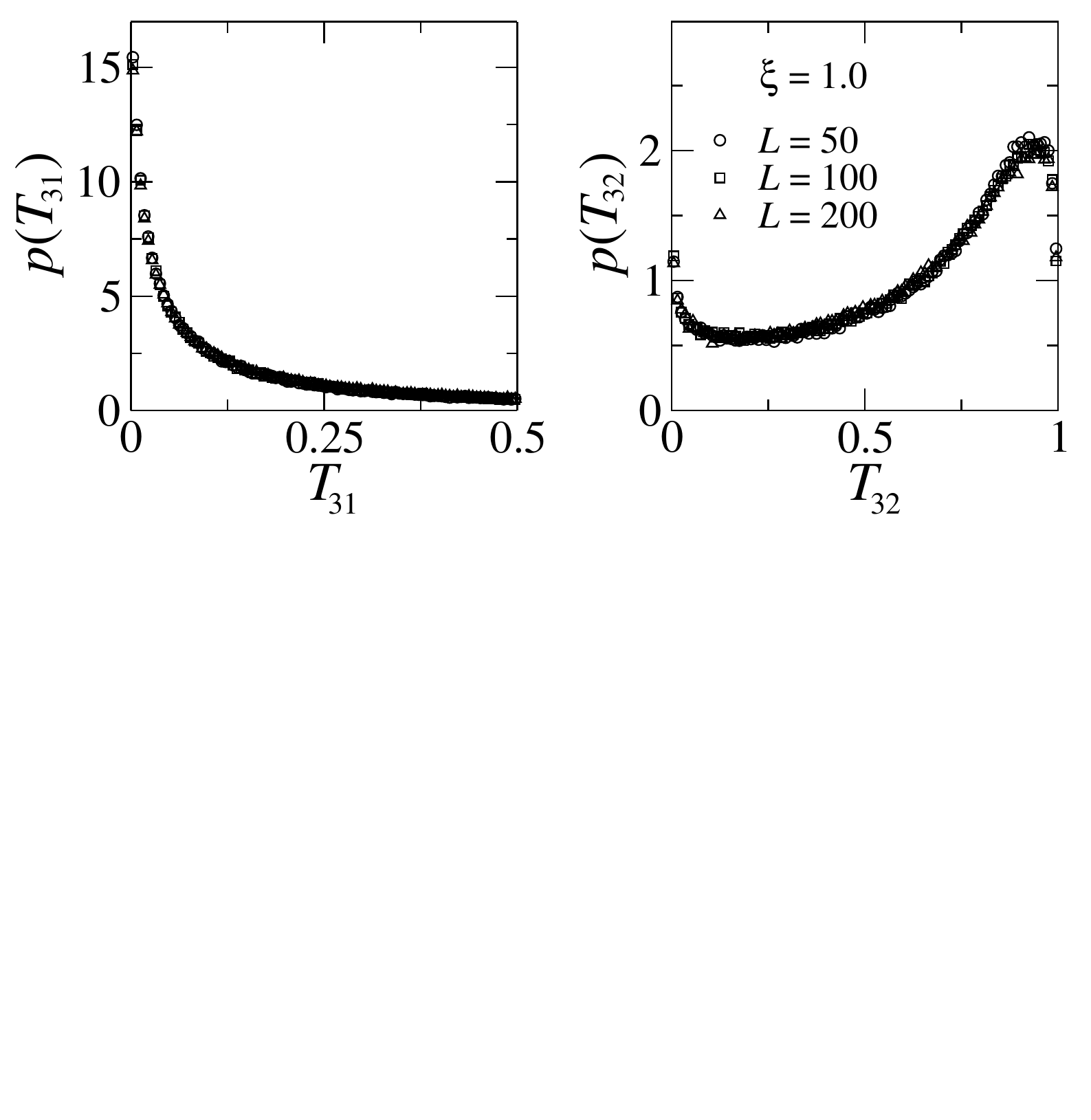}
\caption{{\footnotesize Distribution of $T_{31}$ and $T_{32}$ for a fixed $\xi$ 
and different wire lengths, as indicated in the panels. For the numerical 
calculation we used an ensemble of $2\times 10^{5}$ wire realizations and 100 
bins to construct the histograms.}}
\label{fig:pT3TN}
\end{center}
\end{figure}

From Eq.~(\ref{eq:S1DAM}) we observe that the reflection and transmission 
amplitudes $t$ and $r'$, respectively, that appear in the expression of $f$, 
Eq.~(\ref{eq:f3}), depend on the localization length and the degree of 
disorder. This dependence is only through the ratio $\xi=\ell_{\infty}/L$, from 
which $\xi^{-1}$ (the length of wire in units of the localization length) can be 
considered as the disorder strength, satisfying a single-parameter scaling 
hypothesis.\cite{Anderson1980} This is verified in Fig.~\ref{fig:pT3TN} for the 
distribution of $T_{31}$ and $T_{32}$ for different wire lengths. 

\begin{figure}
\begin{center}
\includegraphics[width=8.5cm]{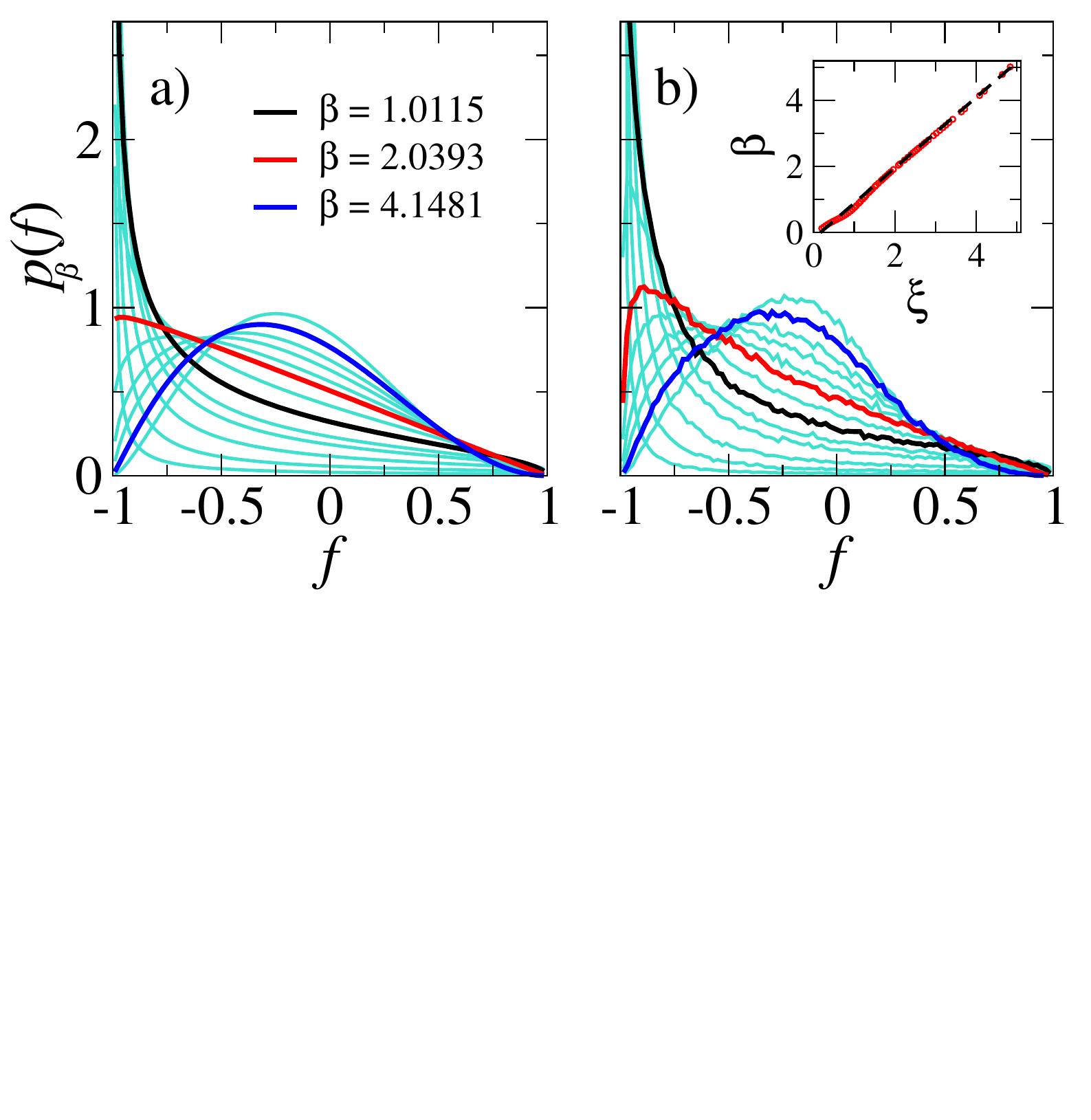}
\caption{{\footnotesize (Color online) Behavior of $p_{\beta}(f)$ for several values of $\beta$ for (a) the analytical expression, Eq.~(\ref{eq:pfAll}), and (b) for numerical simulations of Eq.~(\ref{eq:S1DAM}) with disorder modeled by the 1D Anderson Hamiltonian. The $\beta \approx 1$, 2 and 4 cases, indicated in the inset of panel (a), correspond to the chaotic cases in presence and absence of time-reversal invariance, and presence of symplectic symmetry; respectively. In the inset of panel (b) we show the relationship between $\xi$ and $\beta$ obtained by fitting $p_{\beta}(f)$ to the numerical distributions (see the text).}}
\label{fig:pfBTheory}
\end{center}
\end{figure}

In Fig.~\ref{fig:pfBTheory} we show the behavior of distribution $p_{\beta}(f)$ 
for several values of $\beta$ for (a) the analytical expression, 
Eq.~(\ref{eq:pfAll}), and (b) numerical simulations of $f$ with $r'$ and $t$ 
obtained from Eq.~(\ref{eq:S1DAM}). The cases $\beta\approx 1$, 2, and 4, 
indicated in the inset of panel (a), correspond roughly to the chaotic cases 
in presence and absence of time-reversal invariance, and in presence of 
symplectic symmetry; respectively. The fitting between the analytical 
expression $p_{\beta}(f)$ to the numerical distribution, for each value of the 
ratio $\xi$, determines the corresponding value of $\beta$. We numerically found 
that the parameters $\xi$ and $\beta$ are related through a quadratic equation 
given by
\begin{equation}
\label{eq:BetaFit}
\beta(\xi) \approx -4.449\times 10^{-3}\, \xi^{2} + 1.071\, \xi - 0.1806,
\end{equation}
with a statistical indicator of $\chi^{2}=4.34\times 10^{-4}$, as shown in the 
inset of Fig.~\ref{fig:pfBTheory}~(b). For the fitting we chose $\xi$ in the 
interval [1.7, 3.7] in order to avoid divergencies for values of $f$ close to 
-1. Furthermore, in Fig.~\ref{fig:pfBTheory} we observe some deviations between 
both distributions which are model-dependent, however the phenomenology showed 
by the expression of Eq.~(\ref{eq:pfAll}) is well reproduced. A continuous 
transition between different values of $\beta$ is also observed. For the 
simulations we constructed ensembles of $2\times 10^{5}$ disordered wires and 
used 100 bins to construct the histograms.

In what follows we verify our proposal with a more realistic model of disorder, i.e., with finite-width bulk-disordered waveguides.


\section{Application to bulk-disordered waveguides}
\label{sec:3TMS}

We validate the applicability of our proposal, Eq.~(\ref{eq:pfAll}), by means of 
finite element simulations of bulk-disordered (BD) waveguides. A BD waveguide 
consists of a quasi-one-dimensional wire formed by attaching $N$ two-dimensional 
building blocks (BB). Every building block is a square cavity of side $d$ 
connected to two semi-infinite leads of width $d$ on the left and right sides. 
We place at random a circular obstacle of radius $\rho$ inside each building 
block to produce an ensemble. The leads support plane waves with energy $E$; 
when $E$ lies inside the interval $(\hbar^2/2md^2)\left[\mu^2\pi ^2, \; 
(\mu+1)^2 \pi ^2 \right]$ they support $\mu$ open channels. We use the 
dimensionless units $\hbar^2/2md^2 =1$, so that one open channel (i.e., the case 
we will focus below) occurs for $E \in [\pi^2,(2\pi)^2]$. We fix the energy to 
$E=(1.5\pi)^2$, so that both leads support one open channel and the energy is 
far from the new channel threshold in order to avoid threshold singularities; 
we also set $d=100\rho_0$ with $\rho_0=1$.

To compute the scattering quantities of the bulk-disordered waveguides we use 
the combination rule of scattering and transfer matrices, as shown in 
Ref.~[\onlinecite{AML13}]. First, by means of standard finite element methods (see 
for instance Refs.~[\onlinecite{LMSP02,MLSP02,MLSP03}]) we compute the scattering 
matrix of a $i$-th building block:
\begin{equation}
S_{\mathrm{BB}}^{(i)} = \left(
\begin{array}{cc}
r_i & t'_i \\
t_i & r'_i
\end{array}
\right) \ ,
\end{equation}
where $r_{i}$ $(r'_{i})$ and $t_{i}$ $(t'_{i})$ are the reflection and 
transmission amplitudes, for incidence from the left (right). Then, the 
transfer matrix is easily obtained from the elementary relation with the 
$S$-matrix;\cite{MelloBook} this relation leads to the transfer matrix 
of the building block $M_{BB}^{(i)}$. Therefore, since the building blocks are 
attached in series, the transfer matrix $M$ of the complete waveguide composed 
by $L=N$ building blocks can be easily calculated as
\begin{equation}
M(L) = \prod_{i=1}^{L} M_{\mathrm{BB}}^{(i)} = \left( \begin{array}{cc}
\alpha & \beta \\
\beta^* & \alpha^*
\end{array} \right) \ .
\end{equation}
Finally, the scattering matrix of the waveguide of length $L$ is
\begin{equation}
\label{eq:SBD}
S(L) = \frac{1}{\alpha^*}\left(
\begin{array}{cc}
-\beta^* & 1 \\
1 & \beta
\end{array}
\right) = \left(
\begin{array}{cc}
r & t' \\
t & r'
\end{array}
\right) \ .
\end{equation}

\begin{figure}
\centerline{\includegraphics[width=8.4cm]{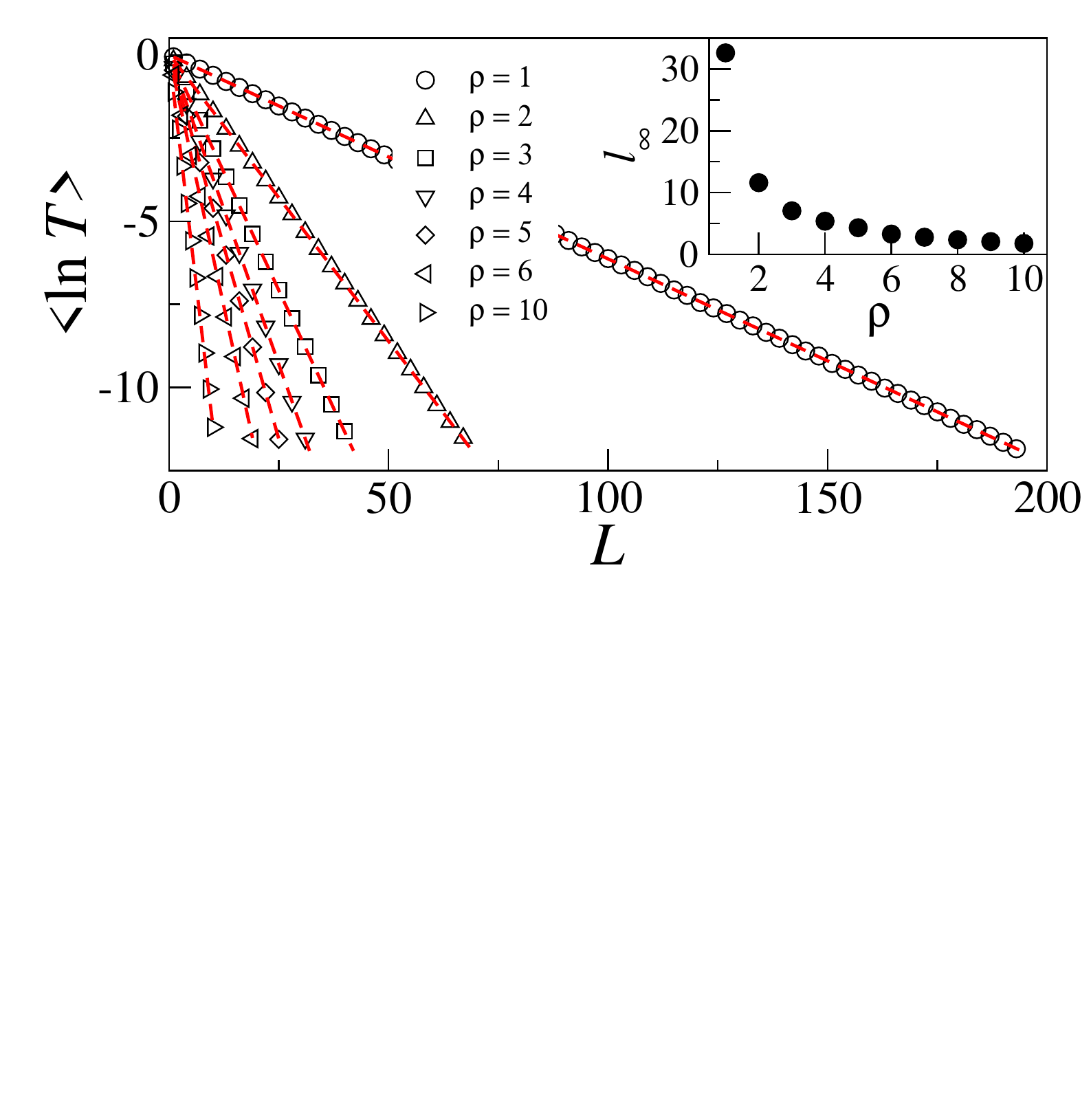}}
\caption{(Color online) Average logarithm of the conductance $\langle \ln g 
\rangle$ as a function of the waveguide length $L$ for bulk-disordered 
waveguides (supporting one open channel) characterized by $\rho=[1,10]$. 
Red-dashed lines are fittings to the data with Eq.~(\ref{ell}); these fittings 
are performed to extract the localization lengths $\ell_{\infty}$. Inset: 
$\ell_{\infty}$ as a function of $\rho$. Each point in the figure is computed 
by averaging over an ensemble of $10^5$ waveguide realizations.}
\label{lng_M1}
\end{figure}

For the statistical analysis we generate an ensemble of bulk-disordered 
waveguides from sets of different building blocks, constructed by randomly 
moving the inner obstacle of radius $\rho$. In Fig.~\ref{lng_M1} we plot the 
average of $\langle\ln T\rangle$, where $T$ is given 
by\cite{Buttiker1986,ButtikerIBM,Landauer1,Landauer2}
\begin{equation}
T(L) = \mathrm{tr}(tt') ,
\end{equation}
as a function of the waveguide length $L$ for bulk-disordered waveguides with 
$\rho=[1,10]$. Notice that the decay of $\langle\ln T\rangle$ vs. $L$ is faster 
the larger the value of $\rho$ is. Thus one can use the radius of the obstacle 
to tune the disorder strength in our waveguides: The larger the value of $\rho$ 
the stronger the disorder strength. Furthermore, by fitting these curves 
to~\cite{Anderson1958}
\begin{equation}
\label{ell}
\langle \ln T \rangle = -2L/\ell_{\infty} = -2/\xi\ ,
\end{equation}
we extract the corresponding localization length $\ell_{\infty}$, see red 
dashed lines in Fig.~\ref{lng_M1}. In the inset of Fig.~\ref{lng_M1} we show the 
obtained values of $\ell_{\infty}$ as a function of $\rho$. They are used to 
design waveguides characterized by specific disorder strengths through the 
ratio $\xi=\ell_{\infty}/L$. We restrict our analysis to building blocks with 
inner obstacles with radius $\rho=1$ to get longer waveguides (see 
Fig.~\ref{lng_M1}), but since the lengths $L$ of the waveguides are given as 
integer multiples of building blocks, not any value of $\xi$ is allowed. 

\begin{figure}
\begin{center}
\includegraphics[width=8.5cm]{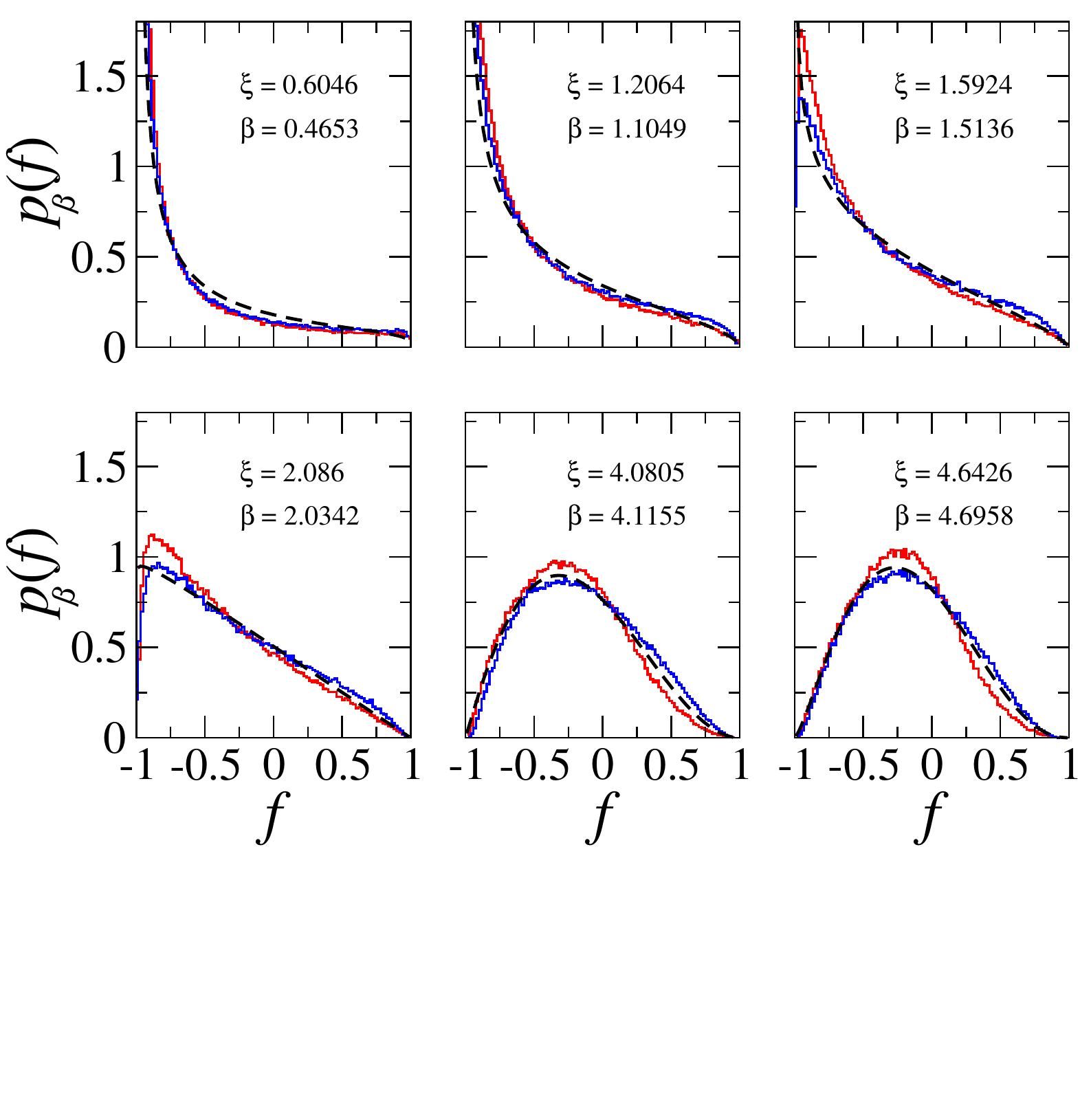}
\caption{{\footnotesize (Color online) Probability distribution, 
$p_{\beta}(f)$, for the three-terminal disordered device. The histograms 
correspond to the distribution of $f$, Eq.~(\ref{eq:f3}), with $r'$ and $t$ 
obtained from numerical simulations for the 1D Anderson model (red) and for 
bulk-disordered waveguides (blue). For the numerical analysis we performed 
ensembles of $2\times 10^{5}$ wire realizations. For the histograms we used 
100 bins. }}
\label{fig:pfXiB}
\end{center}
\end{figure}

In Fig.~\ref{fig:pfXiB} we present probability distributions $p_{\beta}(f)$ for 
the three-terminal disordered device for different values of $\xi$, as 
indicated in the insets. The dashed lines correspond to the analytical 
expression, Eq.~(\ref{eq:pfAll}), with $\beta$ obtained from 
Eq.~(\ref{eq:BetaFit}) for the corresponding $\xi$. The histograms correspond 
to the numerical results obtained from the two models of the disordered wire, 
the 1D Anderson model of Eq.~(\ref{eq:S1DAM}) (red) and for bulk-disordered 
waveguides of Eq.~(\ref{eq:SBD}) (blue). The results show significant 
model-dependent deviations, however the phenomenology captured by 
$p_{\beta}(f)$ is well reproduced for all the transport regimes.


\section{Conclusions}
\label{sec:conclusions}

We studied the voltage drop along a horizontal disordered wire, in a 
three-terminal device. The voltage was measured by means of a third terminal, 
used as a voltage probe, in an asymmetric configuration; that is, when the 
probe is on one side of the wire. Our analysis was based on a random matrix 
theory result accounting for the distribution of the voltage $p_{\beta}(f)$, 
depending only on the Dyson parameter $\beta$ for all symmetry classes:  
orthogonal ($\beta=1$), unitary ($\beta=2$), and symplectic ($\beta=4$). This 
distribution was extended to a continuous parameter $\beta>0$ and proposed as a 
phenomenological expression covering all the transport regimes of the 
disordered wire, from the ballistic to the localized regime. We validate our 
proposal with two models for the disordered wire: The one-dimensional Anderson 
model and bulk-disordered waveguides. It is relevant to stress that the 
parameter $\beta$ in $p_{\beta}(f)$ may be interpreted as the (reciprocal) 
degree of disorder in a wire of length $L$, and characterized by the 
localization length $\ell_{\infty}$, since we found that 
$\beta\approx\ell_{\infty}/L$ in a wide range of disorder strengths. We have to 
admit that our results show significant deviations between the numerical 
distributions and our proposal, however the phenomenology is well reproduced.

It is worth mentioning that given the wide classical wave analogies to 
quantum transport,\cite{Doron1990,Rafael2003,Schanze2005,Kuhl2005,Hemmady2005,
Laurent2006,Bittner2011,Angel2015,Enrique2016} our results can be tested by 
experiments with microwaves or mechanical waves with either surface or bulk 
disordered waveguides.


\acknowledgments

A.M.M.-A. acknowledges Benem\'erita Universidad Aut\'onoma de Puebla (BUAP) and PRODEP under the project DSA/103.5/16/11850 for financial support. JAM-B acknowledges financial support from VIEP-BUAP (Grant No.~MEBJ-EXC18-G), 
Fondo Institucional PIFCA (Grant No.~BUAP-CA-169), and CONACyT (Grant 
No.~CB-2013/220624).  MM-M thanks to CONACyT financial support through the 
Grant No. CB-2016/285776.



\end{document}